\documentclass[superscriptaddress,aps,prd,nofootinbib,twocolumn,showpacs]{revtex4-1}

\usepackage{multirow}
\usepackage{amsmath,epsfig,ulem}
\usepackage{amsfonts}
\usepackage{amsmath}
\usepackage{amssymb}
\usepackage{graphicx}
\usepackage{mathrsfs}
\usepackage{hhline,color}
\usepackage{pifont}
\usepackage{lmodern}
\usepackage[utf8]{inputenc}
\usepackage{ulem}

\def\beq{\begin{equation}}
\def\eeq{\end{equation}}
\def\beqa{\begin{eqnarray}}
\def\eeqa{\end{eqnarray}}
\def\ban{\begin{eqnarray*}}
\def\ean{\end{eqnarray*}}
\def\bi{\begin{itemize}}
\def\ei{\end{itemize}}

\begin{document}

\title{Deconfinement and chiral restoration within the SU(3) Polyakov--Nambu--Jona-Lasinio
and entangled Polyakov--Nambu--Jona-Lasinio models in an external magnetic field}

\author{M\'arcio Ferreira}
\affiliation{Centro de F\'{\i}sica Computacional, Department of Physics,
University of Coimbra, P-3004 - 516  Coimbra, Portugal}
\author{Pedro Costa}
\affiliation{Centro de F\'{\i}sica Computacional, Department of Physics,
University of Coimbra, P-3004 - 516  Coimbra, Portugal}
\author{D\'{e}bora P. Menezes}
\affiliation{Departamento de F\'{\i}sica, CFM, Universidade Federal de Santa Catarina, 
Florian\'opolis, SC, CP 476, CEP 88.040-900, Brazil}
\affiliation{Departamento de F\'isica Aplicada, Universidad de Alicante, 
Ap. Correus 99, E-03080, Alicante, Spain}
\author{Constan\c ca Provid\^encia}
\affiliation{Centro de F\'{\i}sica Computacional, Department of Physics,
University of Coimbra, P-3004 - 516  Coimbra, Portugal}
\author{Norberto N. Scoccola}
\affiliation{Department of Theoretical Physics, Comisi\'on
Nacional de Energ\'{\i}a At\'omica, 1429 Buenos Aires,
Argentina;\\
CONICET, 1033 Buenos Aires, Argentina;
Universidad Favaloro, 1078 Buenos Aires, Argentina.}

\date{\today}

\begin{abstract}
The behavior of the quark condensates at zero chemical potential and finite 
temperature subject to an external magnetic field is studied within the 
three flavor Nambu--Jona-Lasinio model with Polyakov loop (PNJL) and its extension, 
the so-called entangled PNJL model (EPNJL). 
A comparison with recent lattice QCD data is performed and it is shown that at $T=0$ MeV
the light quark condensates are in quantitative agreement. At finite temperature, 
although there is an overall reasonable agreement with several lattice results, it is shown 
that in the lattice calculations the effect due to the electric charge quark difference 
is stronger and  the restoration of the $u$ quark chiral symmetry starts at lower 
temperatures. When considering the EPNJL model with a Polyakov loop scale parameter 
that depends on the magnetic field, it is possible to obtain an earlier rise of the 
Polyakov loop with the increase of the magnetic field and due to the entanglement, 
the inverse magnetic catalysis is found as in the lattice QCD calculations.
\end{abstract}


\maketitle

\section{Introduction}

Understanding matter under extremely intense magnetic fields is one of the 
most interesting topics in modern physics due to its relevance for studies involving 
compact objects like magnetars \cite{duncan}, measurements in heavy ion collisions 
at very high energies \cite{HIC,kharzeev} or the first phases of the Universe \cite{cosmo}.

The structure of the QCD phase diagram in the presence of an external
magnetic field has been subject of several studies
\cite{Klimenko:1991he,Ebert:2003yk,Ferrer:2005vd,Mizher:2010zb,Chatterjee,Chernodub:2011mc}, 
in particular, at zero chemical potential $\mu=0$ (the $T-eB$ plane), see
\cite{baliJHEP2012,ruggieri2,Fraga:2012rr,DElia3} for a review. 
The first analysis about the influence of the magnetic field on the  
chiral-symmetry breaking within the framework of the standard 
Nambu--Jona-Lasinio (NJL) model was made in the late 1980s \cite{Klevansky}.
Recently,  the influence of strong magnetic fields on the 
QCD phase diagram covering the whole $T-\mu$ plane was investigated 
within the SU(3) NJL in the mean field approximation \cite{avancini2012}. 
For finite chemical potentials it is found that the location of the critical end point 
occurs at larger temperatures for stronger fields.

At zero chemical potential, almost all low-energy effective models, 
including the NJL--type models, as well as lattice QCD  (LQCD) calculations
\cite{DElia1,DElia2,Ilgenfritz1,Ilgenfritz2}, found an enhancement of 
the condensate due to the magnetic field (magnetic catalysis)
independently of the temperature.
The magnetic catalysis is the result of a stronger coupling of a quark-antiquark pair 
in the presence of the external magnetic field once the spins of the quarks are 
aligned along the direction of induced magnetic field according to their helicities.
This effect leads to an increase of the transition temperature for chiral 
symmetry restoration as a function of $B$. 
However,  a recent LQCD study \cite{baliJHEP2012,bali2012}, for $N_f=2+1$ flavors 
with physical quarks and pion masses, shows a different behavior in the transition 
temperature, in particular, the suppression of the light condensates 
($u$ and $d$ quarks) by the magnetic field, an effect known as inverse magnetic catalysis. 
This suppression in the crossover region gives a nonmonotonic behavior of 
the condensates as a function of the magnetic field, resulting in a decreasing 
transition temperature  with an increasing magnetic field.

In \cite{endrodi2013}, the reaction of the low-lying Dirac modes to the magnetic field 
was studied, showing that large values of the Polyakov loop are favored by the magnetic field.
Therefore, just as the chiral transition temperature, also the deconfinement transition
temperature is a decreasing function of the magnetic field.
A review of the predictions from low-energy approximations of QCD and previous 
lattice simulations is given in \cite{baliJHEP2012,ruggieri2,Fraga:2012rr,DElia3}.

Calculations of deconfinement and chiral pseudo-critical
temperatures with the SU(2) Polyakov NJL (PNJL) \cite{PNJL}  model and
the entangled PNJL (EPNJL) \cite{EPNJL} influenced by magnetic fields have 
been discussed in \cite{ruggieri2,ruggieri}.
As in almost all other low-energy QCD models, these two models predict that 
the  critical temperature for chiral symmetry restoration increases with the 
increase of an external magnetic field. 
It was also shown that  within the EPNJL that the splitting between the chiral 
and deconfinement transition temperatures is smaller than the splitting 
predicted by the PNJL model \cite{ruggieri2}.

In the present work we analyze the effects of high intensity magnetic fields on 
strongly interacting matter using the SU(3) versions of the PNJL and EPNJL 
models, mainly through the behavior of the light quark condensate and
the chiral and the deconfinement transition temperatures.
Although these models do not describe the inverse magnetic catalysis effect
observed in the lattice calculations of \cite{baliJHEP2012} at the transition 
temperature, some other features such as the behavior of the $u$ and $d$ 
condensates with an external magnetic field $B$ for zero temperature, 
are well reproduced. 
We also show that it is possible to account for the inverse magnetic catalysis 
effect through the parametrization of the Polyakov loop.

This paper is organized as follows. In Sec. II, we present the 
(E)PNJL models used in this work, the Polyakov loop potential, 
and the parameterizations chosen. In Sec. III, the transition temperatures
are calculated as a function of the magnetic field, and results are
compared with LQCD and other effective models. In Sec. IV, the
behavior of the condensates with temperature and the magnetic field
intensity is compared  with the LQCD results.
Then, in Sec. V, we show a possible way of reproducing the inverse magnetic 
catalysis within the EPNJL model.

\section{Model and Formalism}
\label{sec:model}

We describe quark matter subject to strong magnetic fields within the SU(3) PNJL model.
The PNJL Lagrangian is given by \cite{PNJL}:
\begin{eqnarray}
{\cal L} &=& {\bar{\psi}}_f \left[i\gamma_\mu D^{\mu}-
    {\hat m}_f\right ] \psi_f ~+~ {\cal L}_{sym}~+~{\cal L}_{det} \nonumber\\
&+& \mathcal{U}\left(\Phi,\bar\Phi;T\right) - \frac{1}{4}F_{\mu \nu}F^{\mu \nu},
    \label{Pnjl}
\end{eqnarray}
where the quark sector is described by the  SU(3) version of the
NJL model which includes scalar-pseudoscalar and the t'Hooft six fermion interactions that
models the axial $U(1)_A$ symmetry breaking \cite{njlsu3},
with ${\cal L}_{sym}$ and ${\cal L}_{det}$ given by \cite{Buballa:2003qv}:
\begin{eqnarray*}
    {\cal L}_{sym}&=& G \sum_{a=0}^8 \left [({\bar \psi}_f \lambda_ a \psi_f)^2 +
    ({\bar \psi}_f i\gamma_5 \lambda_a \psi_f)^2 \right ] ,\\
    {\cal L}_{det}&=&-K\left\{{\rm det}_f \left [{\bar \psi}_f(1+\gamma_5)\psi_f \right] +
    {\rm det}_f\left [{\bar \psi}_f(1-\gamma_5)\psi_f\right] \right \}
\end{eqnarray*}
where $\psi_f = (u,d,s)^T$ represents a quark field with three flavors,
${\hat m}_c= {\rm diag}_f (m_u,m_d,m_s)$ is the corresponding (current) mass matrix,
$\lambda_0=\sqrt{2/3}I$  where $I$ is the unit matrix in the three flavor space,
and $0<\lambda_a\le 8$ denote the Gell-Mann matrices.
The coupling between the magnetic field $B$ and quarks, and between the
effective gluon field and quarks are implemented  {\it via} the covariant derivative
$D^{\mu}=\partial^\mu - i q_f A_{EM}^{\mu}-i A^\mu$
where $q_f$ represents the quark electric charge  ($q_d = q_s = -q_u/2= -e/3$), 
$A^{EM}_\mu=\delta_{\mu 2} x_1 B$ is a  static and constant magnetic field in the $z$ 
direction and $F_{\mu \nu }=\partial_{\mu }A^{EM}_{\nu }-\partial _{\nu }A^{EM}_{\mu }$.
In the Polyakov gauge and at finite temperature the spatial components 
of the gluon field are neglected:
$A^\mu = \delta^{\mu}_{0}A^0 = - i \delta^{\mu}_{4}A^4$.
The trace of the Polyakov line defined by
$ \Phi = \frac 1 {N_c} {\langle\langle \mathcal{P}\exp i\int_{0}^{\beta}d\tau\,
A_4\left(\vec{x},\tau\right)\ \rangle\rangle}_\beta$
is the Polyakov loop.

To describe the pure gauge sector an effective potential
$\mathcal{U}\left(\Phi,\bar\Phi;T\right)$ is chosen in order to reproduce
the results obtained in lattice calculations \cite{Ratti:2006}:
\begin{eqnarray}
    & &\frac{\mathcal{U}\left(\Phi,\bar\Phi;T\right)}{T^4}
    = -\frac{a\left(T\right)}{2}\bar\Phi \Phi \nonumber\\
    & &
    +\, b(T)\mbox{ln}\left[1-6\bar\Phi \Phi+4(\bar\Phi^3+ \Phi^3)-3(\bar\Phi \Phi)^2\right],
    \label{Ueff}
\end{eqnarray}
where $a\left(T\right)=a_0+a_1\left(\frac{T_0}{T}\right)+a_2\left(\frac{T_0}{T}\right)^2$, 
$b(T)=b_3\left(\frac{T_0}{T}\right)^3$.
The standard choice of the parameters for the effective potential $\mathcal{U}$ is
$a_0 = 3.51$, $a_1 = -2.47$, $a_2 = 15.2$, and $b_3 = -1.75$.

The effective potential exhibits a phase transition from color confinement 
($T<T_0$, $\Phi=0$) to color deconfinement ($T>T_0$, $\Phi \neq 0$), where $T_0$ 
is the critical temperature for the deconfinement phase transition in pure gauge. 
We take $T_0=210$ MeV to account for quark back-reaction. 

Besides the PNJL model, where $G$ denotes the coupling constant of the scalar-type 
four-quark interaction in the NJL sector, we consider an effective vertex
depending on the Polyakov loop ($G(\Phi, \bar{\Phi})$): the EPNJL model.
This effective vertex
\begin{eqnarray}
G(\Phi,
\bar{\Phi})=G\left[1-\alpha_1\Phi\bar{\Phi}-\alpha_2(\Phi^3+\bar{\Phi}^3)
\right].
\label{G}
\end{eqnarray}
generates entanglement interactions between the Polyakov loop and the chiral 
condensate \cite{EPNJL}.
For reasons of consistency we use $T_0 = 210$ MeV also in the EPNJL model.

The parameters of the model, $\Lambda$ a sharp cutoff  in 3-momentum space, 
only for the divergent ultraviolet integrals, the coupling constants $G$ and $K$
and the current quark masses $m_u^0$ and $m_s^0$ are determined  by fitting
$f_\pi$, $m_\pi$ , $m_K$ and $m_{\eta'}$ to their empirical values.
We consider $\Lambda = 602.3$ MeV, $m_u= m_d=5.5$ MeV, $m_s=140.7$ MeV, 
$G \Lambda^2= 1.385$ and $K \Lambda^5=12.36$ as in \cite{Klev_param}.
The parameter set ($\alpha_1, \alpha_2$) must satisfy the triangle region
$\left\{-1.5\alpha_1+0.3<\alpha_2<-0.86\alpha_1+0.32\alpha_2 \right. ,$
$\left. \alpha_2>0\right\},$ with $T_0=150$ MeV.
We choose $\alpha_1=0.25$ and $\alpha_2=0.10$.

The thermodynamical potential for the three flavor quark sector, $\Omega$, 
is written as
\begin{eqnarray}
	\Omega(T,\,B)&=& {\cal U}(\Phi,\bar{\Phi},T) + G(\Phi,\bar{\Phi})\sum_{f=u,\,d,\,s}\left\langle \bar{q}_fq_f\right\rangle^2 \nonumber\\
	&+&4K \, \left\langle \bar{q}_uq_u\right\rangle \left\langle \bar{q}_dq_d\right\rangle
	\left\langle \bar{q}_sq_s\right\rangle  \nonumber\\
	&+&\sum_{f=u,\,d,\,s}\left(\Omega_f^{vac}+\Omega_f^{mag}+\Omega_f^{med}\right),
	\label{Omega}
\end{eqnarray}
where the vacuum $\Omega_f^{vac}$, the magnetic $\Omega_f^{mag}$, the medium contributions
$\Omega_f^{med}$ and the quark condensates $\left\langle \bar{q}_fq_f\right\rangle$ 
have been evaluated with great detail in \cite{prc,hotnjl}. 

To obtain the mean field equations we must minimize the thermodynamical potential (\ref{Omega})
with respect to $\left\langle \bar{q}_fq_f\right\rangle$, $\Phi$ and $\bar{\Phi}$
\cite{prc,hotnjl,costa}.
Finally, according to \cite{bali2012} we define the change of the light condensate 
due to the magnetic field as
\begin{equation}
\Delta\Sigma_{f}(B,T)=\Sigma_{f}(B,T)-\Sigma_{f}(0,T),
\label{Delta}
\end{equation}
with
\begin{equation}
\Sigma_{f}(B,T)=\frac{2m_{f}}{m_\pi^2f_\pi^2}
\left[\left\langle \bar{q}_{f}q_{f}\right\rangle(B,T) \right.
- \left. \left\langle \bar{q}_{f}q_{f}\right\rangle(0,0) \right]+1
\end{equation}
where the factor $m_\pi^2f_\pi^2$ in the denominator contains the pion
mass in the vacuum ($m_\pi=135$ MeV) and (the chiral limit of the) pion
decay constant ($f_\pi=87.9$) MeV in PNJL model.


%


\section{PNJL and EPNJL models in an external magnetic field}
\label{sec:PNJL}

\begin{table}[t]
\begin{center}
      \begin{tabular}{ccccccccccc}
            \hline
   $eB$               & \multicolumn{4}{c}{$\textmd{ PNJL}$} &
                & \multicolumn{4}{c}{$\textmd{ EPNJL}$} \\
 \cline{3-6} \cline{8-11}
  (GeV$^2$)    &$\phantom{m}$   & $T^u_c$ & $T^d_c$ & $T^\chi_c$ & $T^\Phi_c$ &$\phantom{m}$  & $T^u_c$ & $T^d_c$ & $T^\chi_c$ & $T^\Phi_c$ \\
%
        \hline
        0 && 200 & 200 & 200 & 171  &&187 & 187 & 187 &  184             \\
        0.2 && 209   & 208 & 208 & 172 & &193 & 193  & 193 &  187             \\
        0.4 && 226   & 224 & 225 & 174 &&206 & 205  & 206 &  195             \\
        0.6 && 246 & 242 & 244 & 178  && 222 & 221 & 222 &    204        \\
        0.8 && 267   & 257   & 262 & 182& & 240 & 237 & 238 &    214    \\
        1 && 288 & 271   & 279 & 186 && 257 & 252  & 255 &   224     \\
        \hline
    \end{tabular}
    \caption{\label{table:Tc} Pseudo-critical temperatures  in MeV for the chiral transition
    $\left(T^\chi_c=(T^\chi_u+T^\chi_d)/2\right)$ and for the deconfinement ($T^\Phi_c$)
    for both, PNJL and EPNJL, models with $T_0=210$ MeV. 
}
\end{center}
\end{table}

At zero temperature the chiral symmetry of QCD is explicitly broken.
Consequently, at high temperature it is expected that chiral symmetry
be restored.
At $eB = 0$ both, PNJL and EPNJL, models show a crossover transition:
we can only establish a pseudo-critical temperature which depends
on the observable used to define it \cite{costa}.
To identify the pseudo-critical temperature for the chiral transition 
{$T^\chi_c=(T^\chi_u+T^\chi_d)/2$ (being $T^\chi_u$ and $T^\chi_d$
the transition temperatures for $u$ and $d$ quarks, respectively)}
and for the deconfinement ($T^\Phi_c$), we use the location of the peaks for 
the vacuum normalized quark condensates and the Polyakov loop field $\Phi$ 
susceptibilities given, respectively, by
\begin{equation}
C_f = -m_\pi\partial\sigma_f/\partial T,~ 
\sigma_f=\left\langle \bar{q_f}q_f\right\rangle(B,T)/\left\langle \bar{q_f}q_f
\right\rangle(B,0),
\nonumber
\end{equation}
\begin{equation}
C_{\Phi} = m_\pi\partial\Phi/\partial T .
\end{equation}
The multiplication by $m_\pi$ is only to ensure that the susceptibilities are 
dimensionless.

\begin{figure}[t!]
		\includegraphics[width=0.85\linewidth,angle=0]{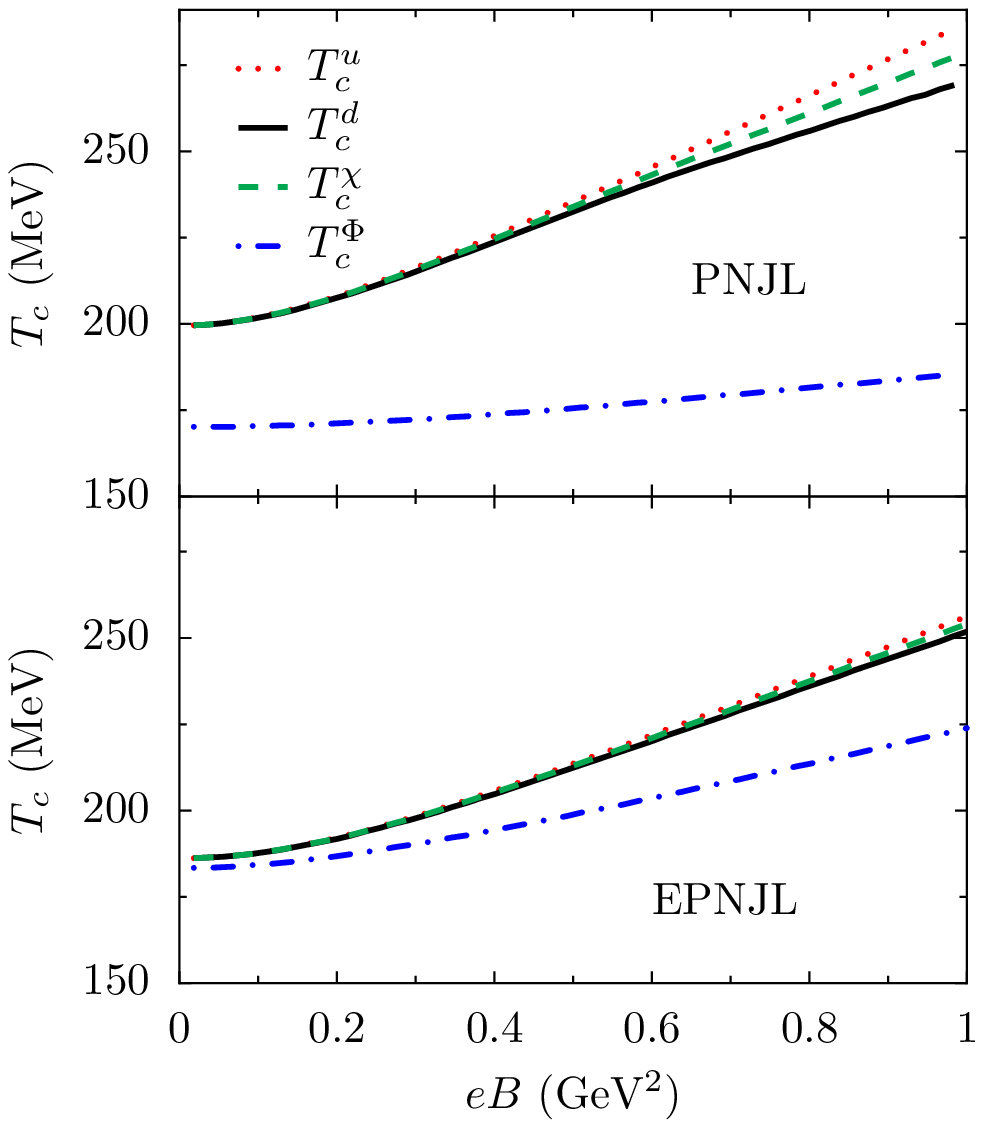}
  	\caption{ 
  	Pseudo-critical temperatures for the chiral transition,
  	$T^\chi_u,\,T^\chi_d,\, \left(T^\chi_c=(T^\chi_u+T^\chi_d)/2\right)$ and for the deconfinement 
  	($T^\Phi_c$) vs the magnetic field intensity for PNJL (top) and EPNJL (bottom).}
		\label{Fig:Temp_crit}
\end{figure}

\begin{figure*}[tb!]
    \includegraphics[width=0.40\linewidth,angle=0]{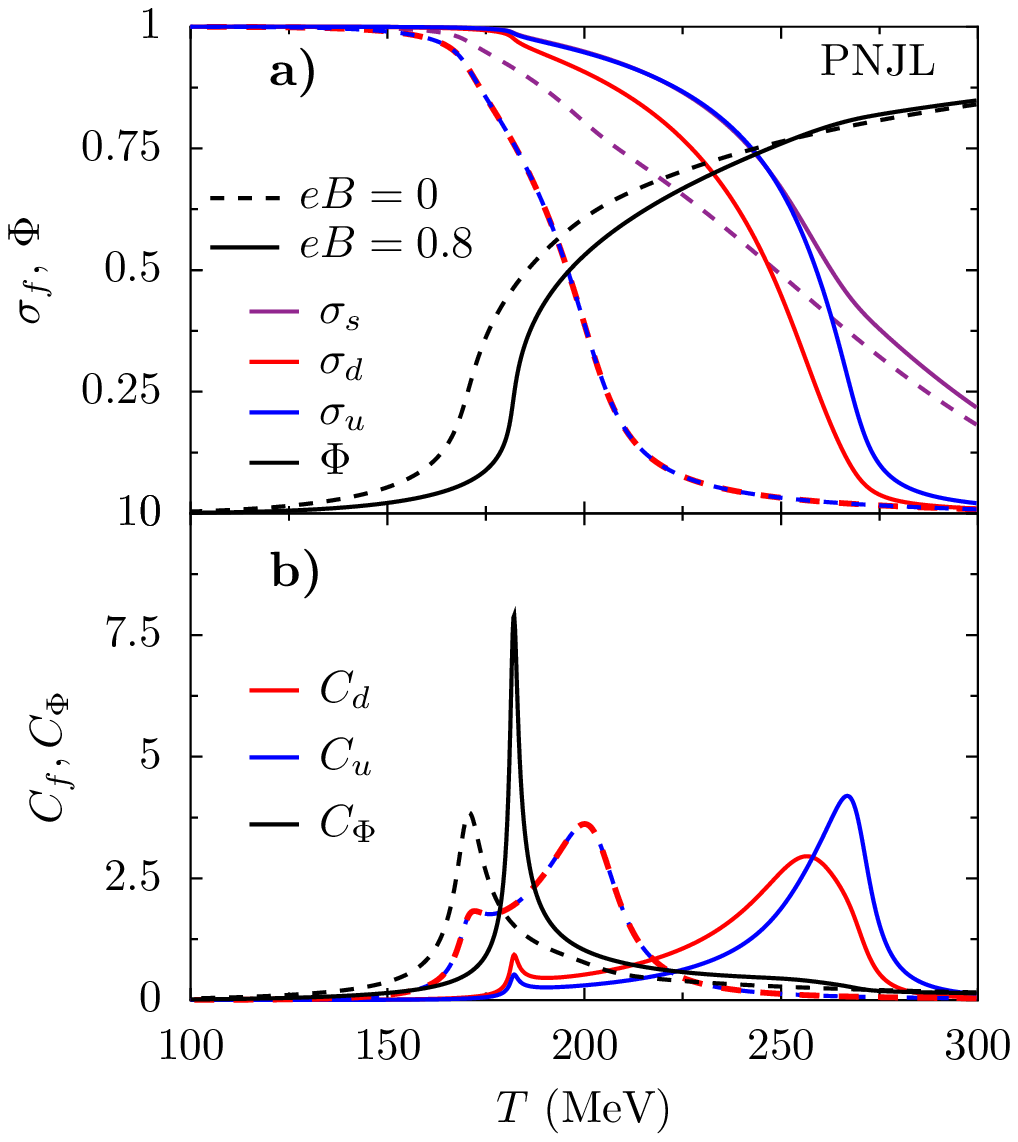} 
    \includegraphics[width=0.40\linewidth,angle=0]{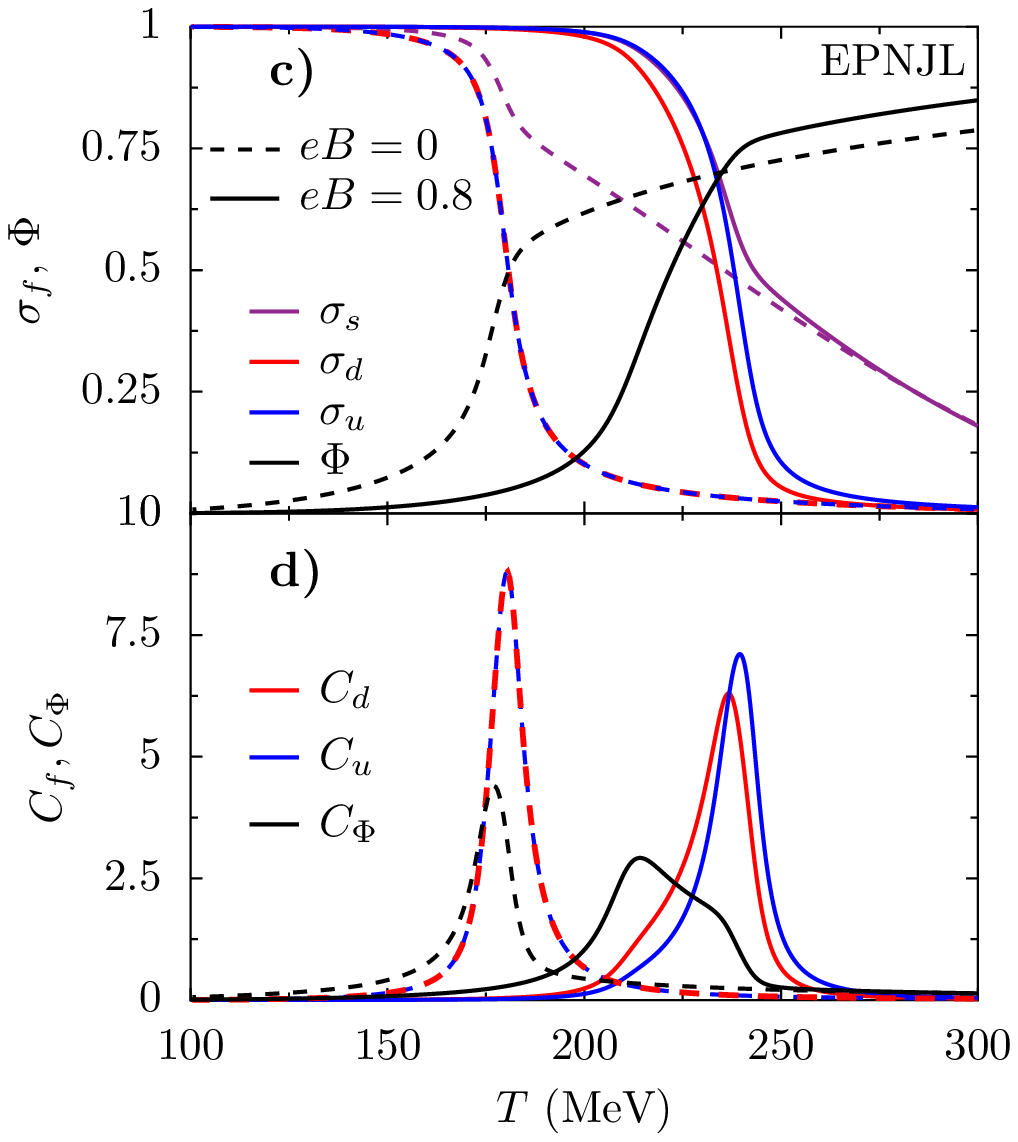} 
    \caption{ 
    Vacuum normalized condensates $\sigma_f$ and the Polyakov loop $\Phi$ (top) and  
    their susceptibilities, $C_f$ and $C_{\Phi}$ (bottom), vs $T$ for  $eB=0$ and 
    $eB=0.8$ GeV$^2$, obtained within  the PNJL (left) and the EPNJL (right).}
    \label{Fig:cond}
\end{figure*}

The pseudo-critical temperatures for $u$ and $d$ quark transitions become different 
as $eB$ increases,  within both PNJL and EPNJL, see Fig. \ref{Fig:Temp_crit} and
Table \ref{table:Tc}, although a stronger difference occurs for PNJL.
Due to its larger electric charge, the $u$ quark has an effective mass that becomes 
larger, which is manifested in the behavior of the respective condensate, 
see Figs. \ref{Fig:cond} (a) and (c) respectively for PNJL and EPNJL,
so the partial restoration of chiral symmetry in the $u$ sector is delayed 
and the respective transition occurs at a higher temperature than the transition in 
the $d$ sector.
It is also observed that as the magnetic field becomes stronger, the 
separation between the temperatures $T^\chi_c$ and $T^\Phi_c$ increases, 
see Figs. \ref{Fig:cond} (b) and (d), and  in Table \ref{table:Tc}. 
This effect is much stronger for the PNJL than the EPNJL, see Fig. \ref{Fig:Temp_crit}. 
In fact, although the entanglement imposed between the quarks and the Polyakov
loop in the EPNJL makes both temperatures $T^\chi_c$ and $T^\Phi_c$ almost
coincident if $eB=0$, a very strong magnetic field destroys this coincidence.

It is also interesting to note from Fig. \ref{Fig:Temp_crit} 
that for both models, we can find a phase in which quark matter is 
(statistically) deconfined, but chiral symmetry is still broken. As pointed 
out in \cite{ruggieri3} this phase can be called constituent quark phase (CQP).
 
In the PNJL,  the magnetic field has a smaller impact on the location of the 
deconfinement crossover as already noticed in \cite{ruggieri} for the SU(2) sector:
$T^\Phi_c$ has just a weak increase, $\sim 15$ MeV if $eB$ increases
from 0 to 1 GeV$^2$.  

This value is comparable with the corresponding 79 MeV increase of the chiral 
transition temperature.
Moreover, the Polyakov loop susceptibilities become narrower with an 
increasing magnetic field and eventually for sufficient strong magnetic fields 
a first order phase transition takes place. 
It is interesting to see a quite different behavior within the EPNJL, where
the deconfinement crossover suffers a shift of 40 MeV , if $eB$ increases
from 0 to 1 GeV$^2$. Due to the entanglement the Polyakov loop
susceptibility peak is shifted towards higher temperatures, together
with the $C_u$ and $C_d$ peaks, which move $\sim68$ MeV, when $eB$ goes
from 0 to 1 GeV$^2$. However, also due to  the entanglement
interaction, the $C_u$ and $C_d$ peaks do not move to so high temperatures
as in the PNJL model.  

The PNJL condensate susceptibilities display small peaks around the peak of 
$C_{\Phi}$ related to the fastening of the phase transition induced by the 
Polyakov loop \cite{costa}. 
They do not signal a phase transition since the variation of the order parameter 
around this temperature is small. A similar effect is seen in the EPNJL Polyakov
loop susceptibility close to the peak of the $C_u$ and $C_d$.

\begin{table}[t]
	\begin{center}
		\begin{tabular}{|c|c|c|c|c|}
    	\hline
      \multirow{2}{*}{$ $}
      & \multicolumn{2}{|c|}{PNJL}
      & \multicolumn{2}{|c|}{EPNJL}
      \\
      \cline{2-5}
      & A & $\alpha$
      & A & $\alpha$
      \\
    	\hline
    	\hline
   		$T_c^u(B)/T_c^u(0)$    & $1.38\times10^{-3}$ & $1.50$ & $6.71\times10^{-4}$ & $1.65$ \\
   		\hline
   		$T_c^d(B)/T_c^d(0)$    & $1.20\times10^{-3}$ & $1.52$ & $5.90\times10^{-4}$ & $1.68$ \\
   		\hline
   		$T_c^\chi(B)/T_c^\chi(0)$    & $1.29\times10^{-3}$ & $1.51$ & $6.31\times10^{-4}$ & $1.67$ \\
   		\hline
   		$T_c^\Phi(B)/T_c^\Phi(0)$    & $5.87\times10^{-5}$ & $1.90$ & $4.42\times10^{-4}$ & $1.61$ \\
   		\hline
		\end{tabular}
	\caption{\label{table:ajuste} Coefficient $A$ and exponent
          $\alpha$ of the expansion of the transition temperatures for
          small values of the magnetic field $eB$, see Eq. (\ref{curvature}).}
	\end{center}
\end{table}

To try to understand the dependence of $T_c^i$ on $eB$ we perform the parametrization 
of the phase transition line introduced in Refs. \cite{DElia1,Skokov}, valid
for small values of the magnetic field ($eB \lesssim 0.5$ GeV$^2$):
\begin{eqnarray}
    \frac{T_c^i(B)}{T_c^i(0)}  = 1 + A \left(\frac{eB}{m_{\pi}^2}\right)^\alpha
    \label{curvature}
\end{eqnarray}
The numerical values of the best-fit coefficients are given in Table \ref{table:ajuste}.
The results show what Fig. \ref{Fig:Temp_crit} also reveals:
the curvature for the Polyakov transition  is softer   in the PNJL model than
in the EPNJL model due to  the entanglement interactions between the
Polyakov loop and the chiral condensate in this last model.

\section{The  PNJL and EPNJL models versus the lattice results}
\label{sec:PNJLvsLattice}

\begin{figure}[t!]
    \includegraphics[width=0.85\linewidth,angle=0]{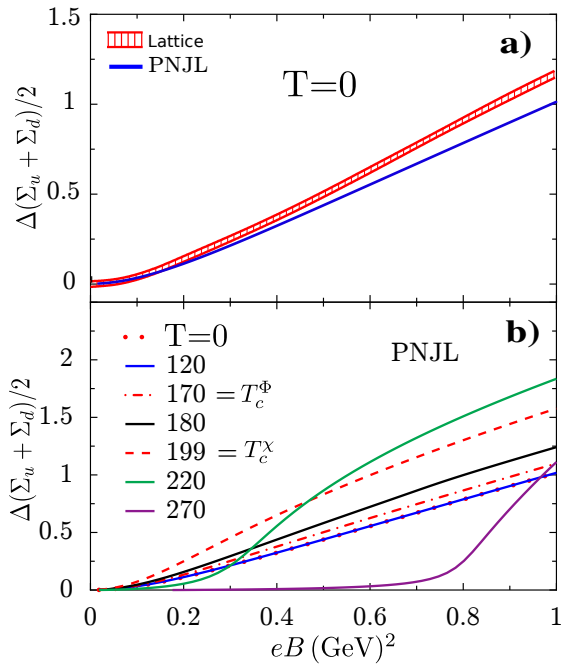}
    \caption{ $\overline{\Delta \Sigma} $ vs $eB$ 
    (a)  at $T=0$ for PNJL  and the lattice results of \cite{bali2012} and
    (b) within PNJL for several temperatures (in MeV) close to the transition temperature. }
\label{Fig:cond_soma}
\end{figure}

Next, we focus our study on the quark condensates as functions of $eB$ at $T = 0$,
having in mind the comparison of the PNJL and EPNJL models with lattice results for 
the quark condensates subject to an external magnetic field \cite{bali2012}.
Note that for $T=0$ the three models NJL, PNJL, and EPNJL coincide.
In Fig. \ref{Fig:cond_soma} (a) we compare the PNJL model results for the change of the
renormalized condensate $\overline{\Delta \Sigma}=\Delta(\Sigma_{u}+\Sigma_{d})/2$
with lattice results extracted from \cite{bali2012}. 
Our results agree quantitatively well and even at $eB = 1$ GeV$^2$, the
discrepancy of the order of $\sim 10\, \%$, is much smaller than the prediction of 
chiral perturbation theory and SU(2) PNJL model (see Ref. \cite{bali2012}).
As expected, for small fields ($eB<m_\pi^2$) we obtain a quadratic dependence of 
$\overline{\Delta \Sigma}$ on $eB$  and a linear dependence for higher fields 
($eB\gg m_\pi^2$) \cite{DElia2}.
In Table \ref{table:vsLatt} we present the lattice results for
the light condensates at zero temperature, as functions of $eB$ \cite{bali2012},
together with the results obtained for the PNJL model.
The average of the light condensates (``$+/2$'') is in very good
agreement with lattice results, especially at low magnetic fields. Even for $eB=1$ GeV$^2$
the average of the light condensates does not differ more than $\sim 10\%$.

\begin{table}[t]
    \begin{center}
{\small
      \begin{tabular}{|c||c|c||c|c||c|c|}
           \hline
            \multirow{2}{*}{$ T=0 $}
            & \multicolumn{2}{|c||}{$eB=0$}
            & \multicolumn{2}{|c||}{$eB=0.2 \textmd{ GeV}^2$}
            & \multicolumn{2}{|c|}{$eB=0.4 \textmd{ GeV}^2$} \\
            \cline{2-7}
            & $+/2$ & $-$
            & $+/2$ & $-$
            & $+/2$ & $-$ \\
            \hline\hline
            (E)PNJL & 1 & 0 & 1.11 & 0.08 & 1.32  & 0.23  \\
            \hline
            Latt. \cite{bali2012} & 1   & 0 & 1.14(2) & 0.09(2) & 1.37(2) & 0.28(2) \\
            \hline\hline
            \multirow{2}{*}{$ T=0 $}
            & \multicolumn{2}{|c||}{$eB=0.6 \textmd{ GeV}^2$}
            & \multicolumn{2}{|c||}{$eB=0.8 \textmd{ GeV}^2$}
            & \multicolumn{2}{|c|}{$eB=1.0 \textmd{ GeV}^2$} \\
            \cline{2-7}
            & $+/2$ & $-$
            & $+/2$ & $-$
            & $+/2$ & $-$ \\
            \hline\hline
            (E)PNJL & 1.55 & 0.40 & 1.79 & 0.58 & 2.02 & 0.76 \\
            \hline
            Latt. \cite{bali2012} & 1.63(3) & 0.47(3) & 1.90(3) & 0.67(3) & 2.16(3) & 0.87(3) \\
            \hline
        \end{tabular} }
        \caption{\label{table:vsLatt} Results obtained for the PNJL (EPNJL) model together with the
        continuum extrapolated lattice results for the light condensates at $T=0$ 
        \cite{bali2012}.
        Columns labeled ``$+/2$'' contain the light condensates average,
        while those with ``$-$'' contain the difference.}
    \end{center}
\end{table}

In Fig. \ref{Fig:cond_soma} (b) the average of $u$ and $d$ condensates is plotted as a 
function of the magnetic field intensity for several temperatures in the PNJL model.
For $T<T^{\chi}_c(eB=0)$ the condensates average increases with $eB$ due to the magnetic 
catalysis effect,  being its value greater the higher the temperature.
When $T>T^{\chi}_c(eB=0)$ we are in the region where the partial restoration of
chiral symmetry already took place. In this region there are two competitive effects:
the partial restoration of chiral symmetry and the magnetic catalysis.
The former effect prevails at lower values of $eB$, making the condensates average
approximately zero. The latter effect becomes dominant as the magnetic field increases
and the average condensate becomes nonzero.
Let us take as an example the case $T = 270$ MeV: since $T = 270$ MeV is larger than
$T^{\chi}_c(eB=0)$ MeV, the average condensate is approximately zero for small values of $eB$ 
and starts to increase around $eB = 0.6$ GeV$^2$, a magnetic field strong enough to prevent 
the restoration of chiral symmetry that would have occurred at zero magnetic field.

The results within the EPNJL model are qualitatively similar to the results of the PNJL model.
However, it is important to make some comments on the new features of EPNJL. 
From Table \ref{table:Tc} it is seen that  the coincidence existing between the 
deconfinement and chiral transition temperatures at $eB=0$ is destroyed in the presence 
of an external magnetic field. When compared with PNJL, the effect of entanglement
present in the EPNJL is seen on the larger (smaller) increase of $T_c^\Phi$  ($T_c^\chi$)
as already explained.

\begin{figure}[tb]
    \includegraphics[width=0.9\linewidth,angle=0]{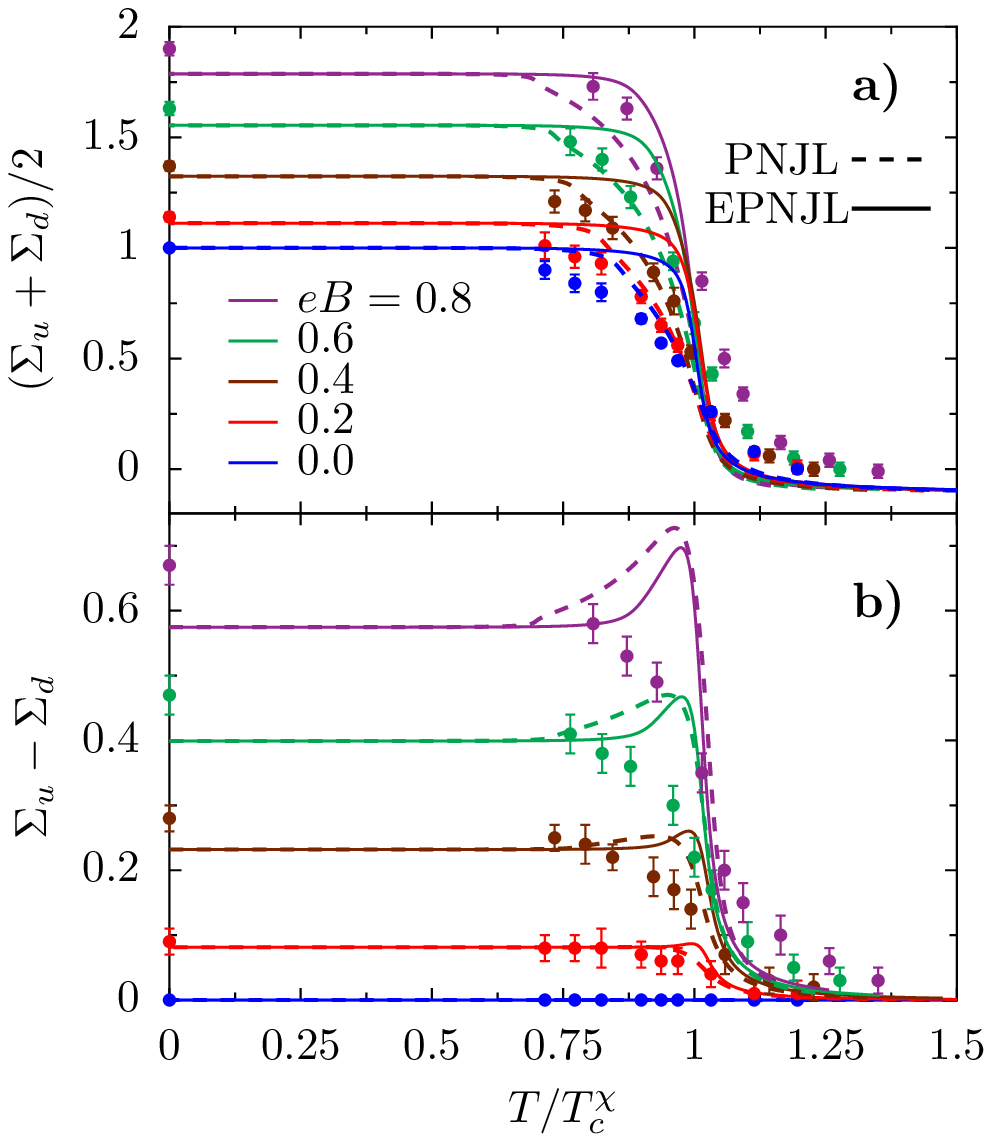} 
    \caption{ 
    (a) Light quark condensate average and 
    (b) light quark condensate difference  and  corresponding lattice results
    taken from \cite{baliJHEP2012,bali2012} vs $T$ for several values
    of $eB$.}
\label{Fig:cond_dif}
\end{figure}

In Table \ref{table:vsLatt} we also list the values for the difference between
$u$ and $d$ quark condensates (``$-$''), at $T=0$, in comparison with lattice calculations.
Once again the results are in good agreement namely for lower values of $eB$, although, a
significative difference between PNJL and lattice calculations occurs for larger values of 
$eB$, with the lattice predicting a larger difference between both condensates.
This means that the effect  due to the electric charge quark difference is
stronger in lattice calculations.

\begin{figure*}[thb]
  \includegraphics[width=0.30\linewidth,angle=0]{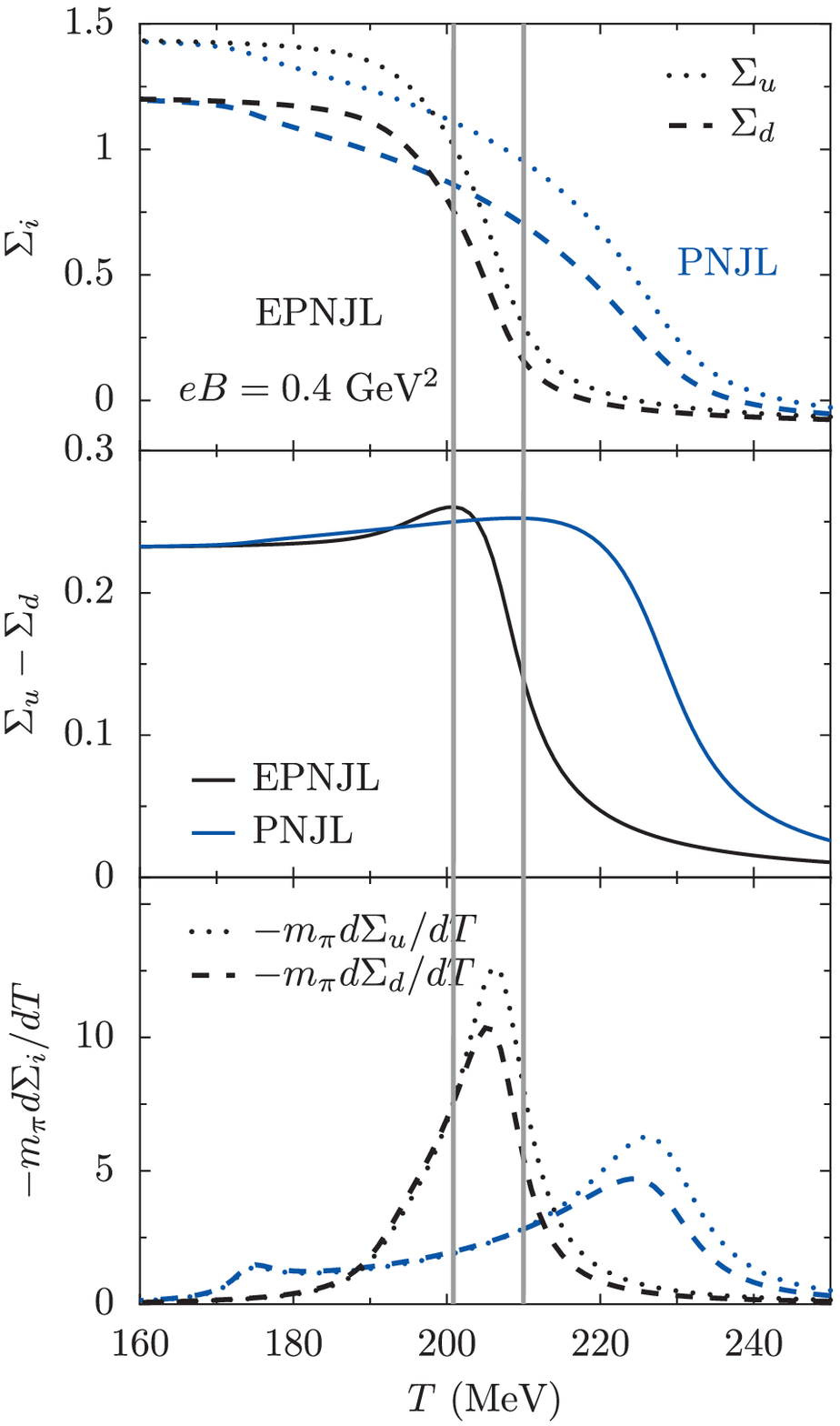}
  \includegraphics[width=0.31\linewidth,angle=0]{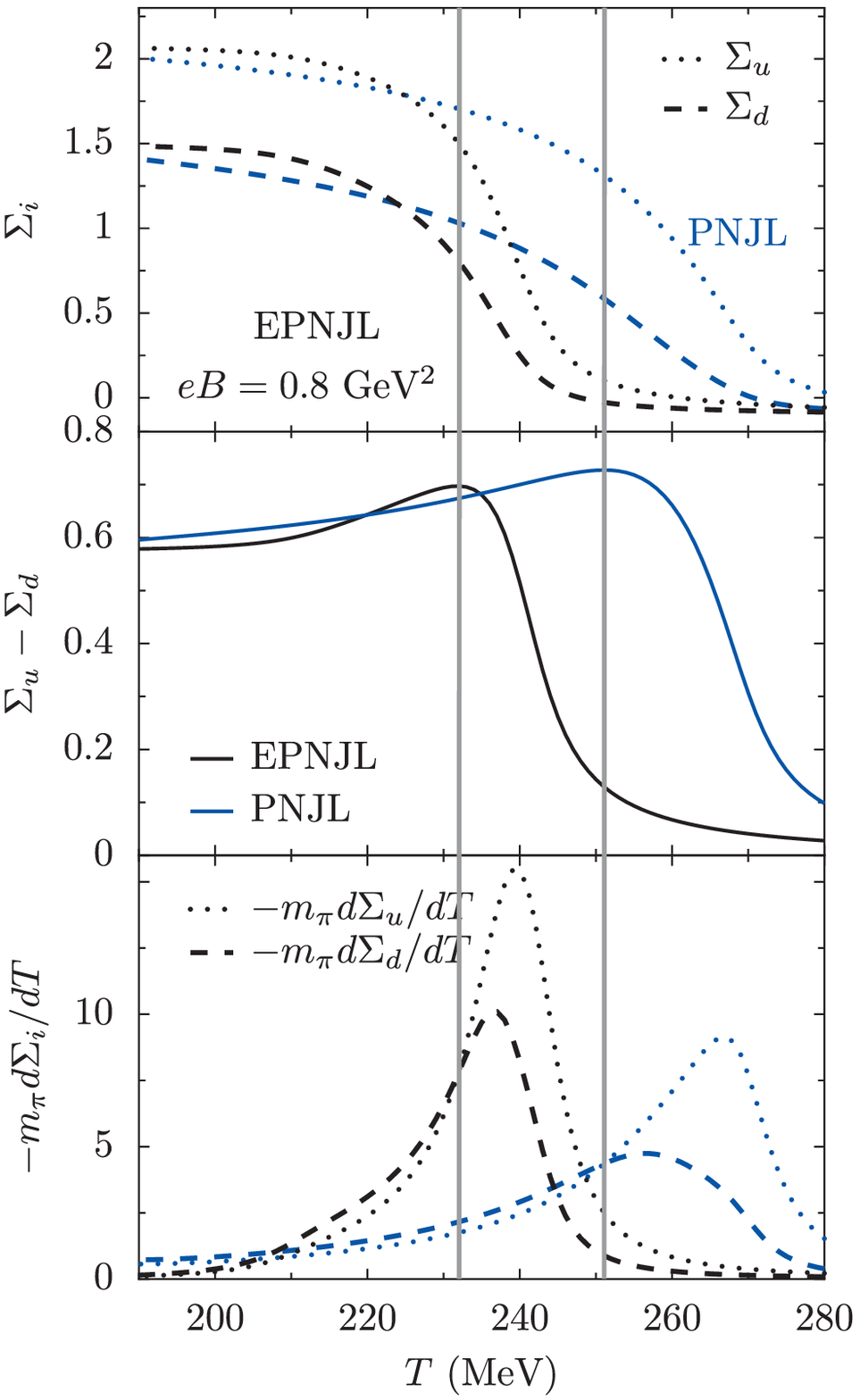}
    \caption{
    The quark condensates $\Sigma_i$ (top), the difference between the
    $u$ and $d$ quark condensates (middle), and derivatives of the
    quark condensates versus $T$ for PNJL (black) and EPNJL (red)
    and two magnetic field intensities, $eB=0.4$ GeV$^2$ (top figures) and
    $eB=0.8$ GeV$^2$ (bottom figures).}
\label{Fig:cond_suscep}
\end{figure*}

In Fig. \ref{Fig:cond_dif} (a) the average  and (b) the difference between light quark 
condensate is plotted as a function of $T/T_c^\chi(eB)$ for several values of $eB$ for PNJL 
(dashed lines), EPNJL (full lines) and lattice data \cite{baliJHEP2012,bali2012}.
The temperature normalization was done in order to remove the inverse magnetic 
catalysis effect in the lattice results. The lattice results for the
condensates and the critical temperatures were taken from \cite{bali2012,baliJHEP2012}.
The comparison plotted in Fig. \ref{Fig:cond_dif} (a) for the average of the light condensates 
shows that in general PNJL and EPNJL have the same behavior as the lattice results 
except for a too fast drop at the respective transition temperatures.
The effect of a stronger magnetic catalysis for the $u$ quark,  due to
its larger electric charge, present in both models at finite temperatures is clear 
in Fig. \ref{Fig:cond_dif} (b):
the larger the magnetic field the larger the difference between $u$
and $d$ condensates, and the respective chiral transition temperatures
(see Table \ref{table:Tc}).
This feature is particularly strong close to the transition temperature, where
the curves for stronger fields have a larger bump. 
This behavior was already found in \cite{Nam} where the authors have employed 
the instanton-liquid model, modified by the Harrington-Shepard caloron solution at finite 
temperature to investigate the chiral restoration in the presence of a strong external
magnetic field.
After the transition temperature $T^\chi_c$, the masses of the quarks are 
smaller, due to the partial restoration of chiral symmetry, prevailing this effect 
over the magnetic catalysis. For these temperatures  the $u$ and $d$ quark condensate 
difference is small.

The bump appears in the $u,\, d$ condensate difference both within PNJL and EPNJL and 
becomes stronger as the magnetic field increases (see Fig. \ref{Fig:cond_dif}). 
To understand the reason of this feature, we show in Fig. \ref{Fig:cond_suscep} 
the condensates $\Sigma_i$, $\Sigma_u-\Sigma_d$, and the $\Sigma_i$ susceptibilities, 
for $eB=0.4$ and $eB=0.8$ GeV$^2$ in both models, removing the temperature renormalization. 
The appearance of the peaks is due to the change of the behavior of the susceptibilities. 
This effect is clearer for $eB=0.8$ GeV$^2$. The vertical gray lines 
indicate the temperature of the $\Sigma_u-\Sigma_d$ maximum. 
For temperatures below this value, $|d\Sigma_d/dT|>|d\Sigma_u/dT|$, and above the bump 
the opposite happens. Due to the charge difference, the magnetic catalysis is stronger 
for $u$ than $d$ quarks, therefore
(a) at lower temperatures, the decrease of the $d$ condensate with temperature is faster, 
because the partial restoration of chiral symmetry in the $u$ sector is delayed; 
(b) at  temperatures close to the transition temperature, $\Sigma_u$  must decrease with 
temperature faster than the $\Sigma_d$. 
Therefore, $\Sigma_u-\Sigma_d$ remains constant at low temperatures, then, it increases 
up to a value below the $d$  chiral transition temperature, and finally, decreases until 
the chiral symmetry is restored.
At variance, the lattice results \cite{bali2012}, predict a monotonous decrease of 
$\Sigma_u-\Sigma_d$ with $T$, possibly showing that  the partial
restoration of chiral symmetry in both the $u$ and $d$ sector occur 
simultaneously.

\begin{figure}[thb]
    \includegraphics[width=0.85\linewidth,angle=0]{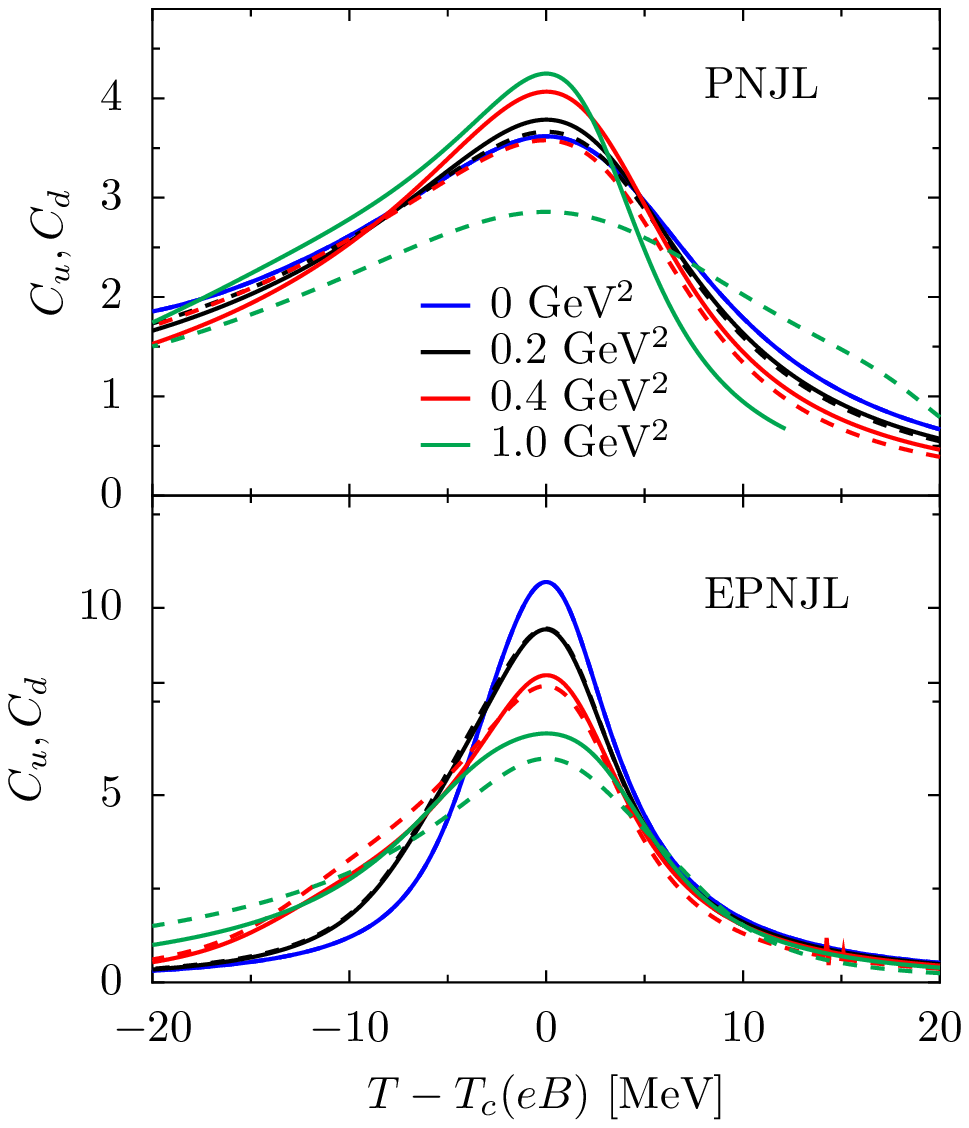}
    \caption{
    Relative changes in the temperature dependence of 
    chiral susceptibility for quark $u$ (full lines) and $d$ (dashed lines) 
    and for different values of $eB$, within PNJL (top) and EPNJL (bottom). 
    The width increases with $eB$.}
\label{Fig:suscep}
\end{figure}

Lattice results \cite{baliJHEP2012} show that the transition remains an analytic 
crossover for magnetic fields at least up to $1$ GeV$^2$. 
In particular, the $u$ quark transition width decreases only mildly and the height 
grows significantly. 
In Fig. \ref{Fig:suscep} the susceptibilities $C_u$ and $C_d$ are plotted  
as a function of $T_c-T_c(eB)$ for several magnetic field intensities. 
In both models the widths of these susceptibilities do not 
change much with the magnetic field intensity even if they show opposite behaviors, 
which can be seen by taking the width of the peak at half maximum.
In the PNJL model for the $u$ quark we see that there is a decrease 
from $\sim 31.1$ to $\sim 21.7$ MeV as the external field is increased 
from zero to $eB = 0.4$ GeV$^2$ and then it remains almost constant. 
The $eB = 0$ width of the peak at half maximum of $C_u$ and $C_d$ is large due to 
the influence of the deconfinement transition which can be seen in 
Fig. \ref{Fig:cond} (b) (red and blue dashed lines).
For the $d$ quark we have a similar behavior except for $eB\approx1$ GeV$^2$ 
where there is a strong increase of the width due to the influence of the 
$u$ quark transition on the $d$ quark transition. The influence of the
$d$ quark transition on the $u$ quark transition also exists but it is very weak. 
Both behaviors are due to the split of the respective transitions as $eB$ increases.    

Contrary to what happens in the PNJL model, the behavior for the EPNJL model shows 
an increase of the widths from $\sim 8.5 (8.5)$ to $\sim 14.7 (16.7)$ MeV as 
the external field increases from zero to $eB = 1$ GeV$^2$ for the $u(d)$ quarks. 
This behavior is mainly explained by the separation of the deconfinement and chiral transitions 
and the influence of the first one on the second one due to the entanglement interaction.
The height also presents different behaviors in both models, and different from LQCD:
in the PNJL  the height increases $\sim18\%$ for $C_u$ and  decreases $\sim21\%$
for $C_d$, for $eB$ between 0 and $1$ GeV$^2$. Within the EPNJL, $C_u$ decreases 
$\sim38\%$ from $eB=0$ to $1$ GeV$^2$. 
This result is a consequence of the entanglement with the Polyakov loop 
(see Fig. \ref{Fig:cond}).

Due to the charge difference, the magnetic field softens more the $d$ transition
than  the $u$ transition once the peaks height is always bigger for the $u$ quark in 
both models.

\section{Inverse magnetic catalysis at finite $T$}
\label{sec:inverse}

So far, we have seen that SU(3) (E)PNJL reproduces quite well
the lattice QCD quark condensate behavior in an external magnetic field
except for the inverse magnetic catalysis effect predicted by lattice QCD
calculations at temperatures of the order of the transition temperature
and  high magnetic fields, and the width of the susceptibilities at
the transition. In the (E)PNJL the deconfinement is described by the Polyakov 
loop which couples weakly to the magnetic field as referred above.
It should be noted that the Polyakov loop potential was originally parametrized
in order to reproduce the pure gluonic lattice data. 

Later, it was realized that the inclusion of dynamical quarks leads to a decrease
of the scale parameter $T_0$. Since strong magnetic fields have certainly an effect
on dynamical quarks, one expects that their presence could affect the value of $T_0$. 
In a recent lattice calculation \cite{endrodi2013}, it is argued that the inverse 
magnetic catalysis may be a consequence of how the gluonic sector reacts to the presence 
of a magnetic field: the distribution of gluon fields may change as an indirect 
effect of the magnetic field mediated by quark loops that destroys the chiral 
condensate. This behavior happens around and above the deconfinement temperature 
differently from lattice results at zero temperature \cite{DElia2}, where the modified
distribution of gluon fields contributes to increase the magnetic catalysis.
The authors have shown that the change of the renormalized Polyakov loop increases 
sharply with the magnetic field around the transition temperature, and that the 
transition temperature decreases with the magnetic field.
Therefore, the back-reaction of the quarks on the gauge fields should be 
incorporated in effective models in order to describe the inverse
magnetic catalysis.

On the other hand, it is known the effect of screening of the gluon interactions  
in a magnetic field in the region of momenta relevant for the chiral symmetry 
breaking dynamics \cite{Miransky}. 
In this region, gluons acquire a mass $M_g$ of order $\sqrt{N_f \alpha_s|eB|}$, with
$M_g$ being the mass of a quark-antiquark composite state coupled to the gluon field. 
In a strong enough magnetic field, this mass $M_g$ for gluons becomes larger. 
This, along with the property of the asymptotic freedom ($\alpha_s$ decreases with 
increasing $eB$), leads to the suppression of the chiral condensate. 
It was also shown that the confinement scale in the presence of a strong magnetic 
field is much less than the corresponding scale in QCD without magnetic field \cite{Miransky}.
Also calculations in the large $N_c$ limit of the quark mass gap in a magnetic field showed
that the quark mass gap does not grow much beyond $\Lambda_{QCD}$. In this scenario,
the screening effects for gluons were not explicitly taken into account.
By introducing $1/N_c$ corrections the screening effects should grow as $eB$ increases
leading to the reduction of the pseudo-critical temperatures \cite{Kojo:2012js}.

One possible approach to mimic the reaction of the gluon sector to the presence of an external
magnetic field is to choose a magnetic field dependent $T_0(eB)$, in order to reproduce the
correct transition temperatures given by lattice \cite{baliJHEP2012,endrodi2013}. 
This type of procedure on $T_0$ had already been proposed in \cite{wambach} in a different 
context: based on renormalization group arguments, an explicit quark chemical potential 
and $N_f$ dependence on $T_0$ in the Polyakov loop potential takes into account 
the back-reaction of the quark degrees of freedom on the Polyakov loop.  

We next start from the lattice results and analyze whether they can be
reproduced  within the PNJL and/or EPNJL models imposing a
dependence of the Polyakov loop on the magnetic field. This dependence
will be included through the parameter $T_0$. However, it should be
pointed out that a too small value of $T_0$ leads to a first order
phase transition within PNJL and EPNJL, and, therefore, the range of
$T_0$ values of interest is limited to the values that maintain the
crossover transition. 

Within the PNJL it is not possible to implement the above scheme because the chiral 
transition temperatures increase strongly with the external magnetic field. 
In order to bring these temperatures down it would be necessary to use very small 
values of $T_0$, for which the deconfinement phase transition becomes of first order. 
However, within EPNJL the chiral condensates and the Polyakov loop are entangled. 
Thus, the chiral transition temperatures are pulled down to temperatures close to the 
deconfinement transition temperature. This model, however, still predicts a first order 
transition for both transitions when $T_0$ is too small at moderate magnetic fields. 

In order to proceed, we take a magnetic field dependent $T_0(eB)$ of the form
\begin{equation}
T_0(eB) = T_0(eB = 0) + \zeta (eB)^2 + \xi (eB)^4,
\label{t0b}
\end{equation}
fitted to the transition temperature for the strange quark number susceptibility data,
that is viewed as a quantity signaling the deconfinement transition,
extracted from \cite{baliJHEP2012}. 

\begin{figure}[t!]
    \includegraphics[width=0.85\linewidth,angle=0]{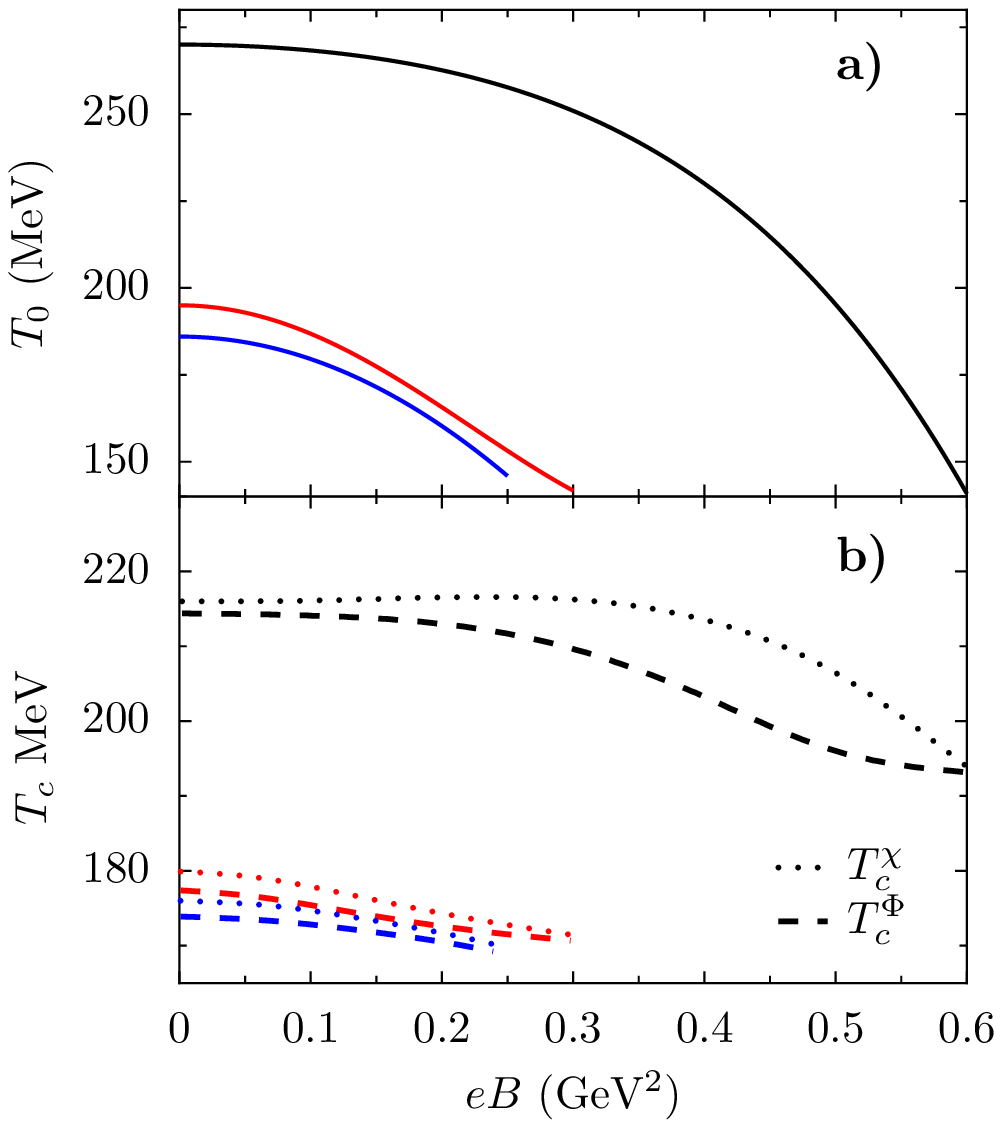}
    \caption{
    {
    (a) $T_0$ as a function of $eB$ defined by Eq. (\ref{t0b}) for the different values of 
    $T_0(eB=0)$ presented in Table \ref{table:T0(eB)} and 
    (b) the corresponding pseudo-critical temperatures as a function of $eB$ for  
    different values of $T_0(eB)$.
    }}
\label{Fig:T0_EB}
\end{figure}

For $eB=0$ we have $T_0 =186$ MeV, in agreement with the
$T_0$ value that encodes the back-reaction of the matter sector to the
gluon sector for $N_f=2+1$ massless flavors \cite{Pawlowski2011}.
The respective transition temperatures are $T^\Phi_c = 173.9$ and $T^\chi_c = 176.0$ MeV 
(see Table \ref{table:T0(eB)}).
\begin{table}[t]
\begin{center}
    \begin{tabular}{cccccc}
        \hline
        $T_0 (eB=0)$	& 	$T^\Phi_c$	&		$T^\chi_c$	&		$eB^{max}$			&	$\zeta$	&	$\xi$ \\
        
        [MeV]    			& 	[MeV]     	& 	[MeV]     	& 	[GeV$^2$]				&[MeV/GeV$^4$]	& [MeV/GeV$^8$]     \\
        \hline
        186			& { 173.9}	 			& 176.0 			& 0.25 				& $-646.491$  &78.8961 \\
        195			& { 177.4}	 			& 179.9 			& 0.3  				& $-845.467$ &2813.38\\
        270			& 214.0				& 216.0 			& 0.61 				& $ -162.632$  &$-545.027$ \\
        \hline	
   \end{tabular}
    \caption{\label{table:T0(eB)} {Pseudo-critical temperatures for chiral transition
    and for the deconfinement in the EPNJL model for different values of $T_0(eB=0)$.}}
\end{center}
\end{table}
The values of $\zeta$ and $\xi$ in Eq. (\ref{t0b}) are also given in Table \ref{table:T0(eB)}.
This parametrization of $T_0(eB)$ can lead to  inverse magnetic
catalysis,  see Fig. \ref{Fig:T0_EB} blue line,  and allows to 
describe the back-reaction on the Polyakov loop due to the presence of an external 
magnetic field for $eB \lesssim  0.25$ GeV$^2$. 
Above this value a first order phase transition is obtained.
A similar scenario also occurs if $T_0(eB)$ is fitted to reproduce the
upper limit of the deconfinement transition shown in Fig. 10 of \cite{baliJHEP2012}.
At $eB=0$ we have $T_0 =195$ MeV with $T^\Phi_c = 177.4$ MeV and $T^\chi_c = 179.9$ MeV.
This parametrization is valid for $eB \lesssim  0.3$ GeV$^2$ (see Table \ref{table:T0(eB)}
and Fig. \ref{Fig:T0_EB}, red line).

A larger range of validity would have been obtained if the quark
back-reaction had not been accounted for (Fig. \ref{Fig:T0_EB}, black lines).
In this case we would have for $eB = 0$, $T_0 = 270$ MeV as obtained in pure gauge, 
which gives $T^\Phi_c = 214$ MeV, 40 MeV higher than the prediction of lattice QCD 
data in \cite{baliJHEP2012}. 
This parametrization also shown in Table \ref{table:T0(eB)},
reproduces lattice QCD data  for $T_c^{\Phi}(eB)$ \cite{baliJHEP2012}, 
shifted by an amount of 40 MeV, for magnetic fields up to $0.61\,\mbox{GeV}^2$.
Above  0.61 GeV$^2$, a  first order phase transition is obtained. 
We next use the last scenario to illustrate our results because larger magnetic 
fields are achieved.

\begin{figure*}[t!]
    \includegraphics[width=0.45\linewidth,angle=0]{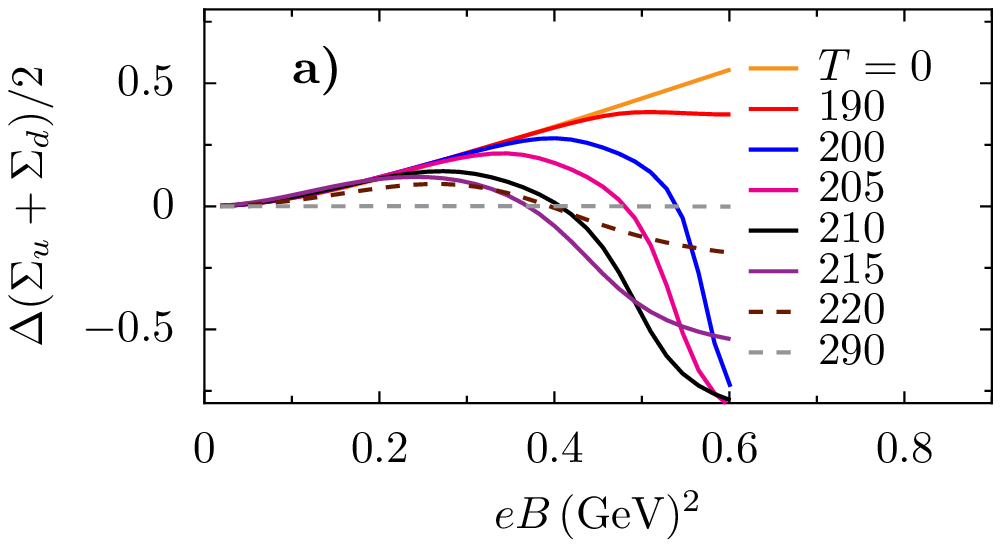}
    \includegraphics[width=0.45\linewidth,angle=0]{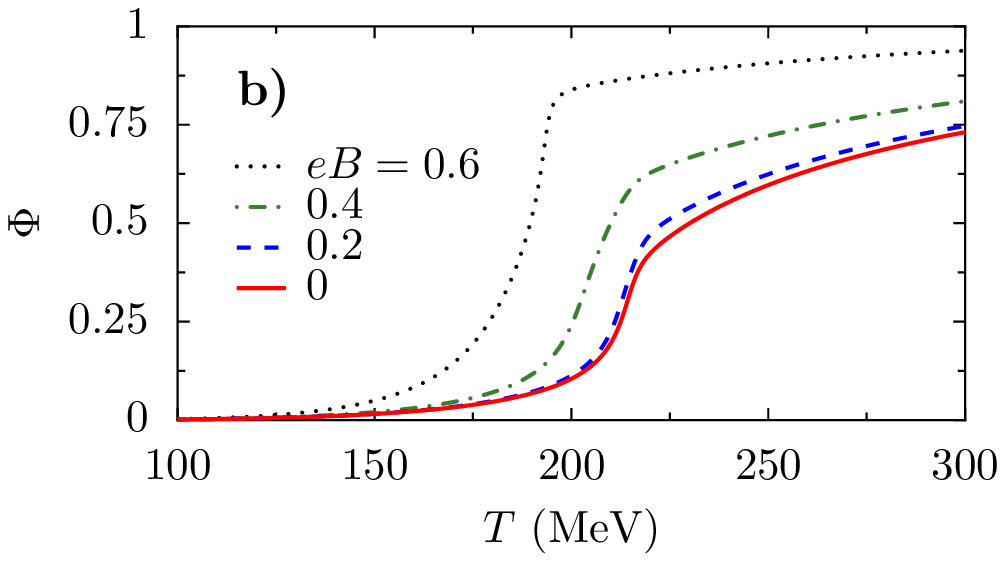}
    \caption{ 
    (a) $\overline{\Delta\Sigma}$ vs $eB$  GeV$^2$  for several
    temperatures (in MeV)  within the EPNJL model with  $T_0(eB)$ 
    defined in Eq. (\ref{t0b})  and 
    (b) the Polyakov loop vs $T$ for different values of $eB$ (in GeV$^2$).
    }
\label{Fig:bali_EPNJL}
\end{figure*}

In  Fig. \ref{Fig:bali_EPNJL} (a), $\overline{\Delta \Sigma}$ is plotted
as a function of the magnetic field for $eB<0.61$ GeV$^2$ and several 
temperatures close to $T_c^{\Phi}(eB=0)$. The main conclusions are:
\textbf{(a)} the  qualitative behavior shown in Fig. 2 of Ref. \cite{bali2012}
and in Fig. 6 of Ref. \cite{endrodi2013} are reproduced, that is, 
the nonmonotonic behavior of the condensates as a function of the magnetic field;
\textbf{(b)} the $T=0$ curve has the highest $\overline{\Delta \Sigma}$,
contrary to the results of Fig. \ref{Fig:cond_soma} (b)  for PNJL with fixed $T_0$;
\textbf{(c)} for $200<T<220$ MeV the strong interplay between the partial
restoration of chiral symmetry, that for stronger magnetic fields occurs at smaller
temperatures (see Table \ref{table:Tc}), gives rise to curves that for small $eB$ values
increase (magnetic catalysis) and as soon as the partial restoration of chiral
symmetry becomes dominant the curve starts to decrease;
\textbf{(d)} for temperatures higher than 190 MeV, the curves are negative because
$\Delta\Sigma(B,T)$ includes the subtraction of the condensate $\Sigma(0,T)$, when  partial
restoration of chiral symmetry has not occurred yet; 
\textbf{(e)} the behavior of the Polyakov loop shown in Fig. \ref{Fig:bali_EPNJL} (b) follows the 
same tendency predicted by the lattice calculations \cite{endrodi2013} and increases with $eB$ 
for a given $T$. Similar results are obtained  if the zero magnetic field  Polyakov loop scale
$T_0$ includes quark back-reaction, however in a smaller range of the magnetic field intensity
if we restrict results to a crossover.

{
To understand the inverse magnetic catalysis phenomenon several studies have been made.
The magnetic inhibition can be a feasible explanation for the decreasing behavior of the chiral
restoration temperature with increasing $eB$ \cite{Fukushima:2012kc}.
Also recently, a mechanism to explain the inverse magnetic catalysis around the 
critical temperature as induced by sphalerons was proposed \cite{Chao:2013qpa}.
}

\section{Conclusions}
\label{sec:conclusions}

In the  present work the behavior of the quark condensates at zero chemical potential
and finite temperature under the influence of an external magnetic field are studied within
three flavor PNJL and EPNJL.
The results are compared with the lattice QCD data discussed in \cite{baliJHEP2012,bali2012}.

Most of the properties of the quark condensates obtained with the 3 flavor version of PNJL
and EPNJL had been obtained with the two flavor versions \cite{ruggieri,ruggieri2}.
In particular, in the present work we have shown that the chiral and deconfinement transition 
temperatures increase in the presence of an external  magnetic field, although the deconfinement 
transition temperature suffers a much weaker effect. Moreover, it was shown  within the SU(3) PNJL 
and EPNJL models that at $T=0$ the quantitative behavior of light quark condensates with the magnetic
field is closer to the lattice results. 
 
Another aspect that should be referred is the effect of the magnetic field on the EPNJL 
deconfinement and chiral transition temperatures: 
the existing coincidence at $eB=0$ is destroyed by the magnetic field. 
Also, the chiral and the deconfinement transition temperatures behave differently with 
the magnetic field in both models: 
the deconfinement temperature suffers just a small increase compared to the huge increase 
of the chiral transition temperature within PNJL, while the increase of the deconfinement 
phase transition is almost three times as large in the EPNJL due to the entanglement interaction. 

The light quark condensates are in good agreement with the  LQCD results, in particular, 
for lower values of $eB$, although, a significative difference between PNJL and lattice 
calculations occurs for larger values of $eB$.
The lattice predicts a larger difference between both condensates at temperatures well below 
the transition temperature. Close to the transition temperature the lattice predicts a softer 
restoration of the chiral symmetry. These two features seem to indicate, that compared with PNJL 
and EPNJL in the lattice calculations, the effect due to the electric charge quark difference 
is stronger and  the restoration of the $u$ quark chiral symmetry starts at lower temperatures. 
The light quarks susceptibility width does not suffer a large effect with $eB$ just like in LQCD, 
while the height of the $u$ quark susceptibility slightly increases (PNJL) or even decreases 
(EPNJL) while in LQCD it suffers a large increase if $eB$ increases from 0 to 1 GeV$^2$.

The magnetic field back-reaction on the Polyakov loop may be taken into account using a magnetic 
field dependent scale parameter $T_0$ which reproduces the lattice transition temperatures. 
When this is done the behavior of the Polyakov loop with $eB$ follows the lattice trend. 
However, within EPNJL a first order phase transition, instead of a crossover, is obtained above 
$\sim 0.3$ GeV$^2$ ($\sim 0.61$ GeV$^2$) taking (not taking) into account the quark back-reaction 
on the Polyakov loop. Understanding the origin of these effects needs further investigation.

\vspace{0.25cm}
{\bf Acknowledgments}:
This work was partially supported by Project No. PTDC/FIS/ 113292/2009
developed under the initiative QREN financed by the UE/FEDER through the
program COMPETE $-$ ``Programa Operacional Factores de
Competitividade'', by Grant No. SFRH/BD/51717/2011, by CNPq/Brazil and
FAPESC/Brazil, by CONICET (Argentina) under Grant No. PIP 00682,
and by ANPCyT (Argentina) under grant PICT11-03-113. We thank fruitful
discussions with Dr. M. B. Pinto and the use of one preliminar code
developed by Dr. S. S. Avancini.
We are also grateful to F. Bruckmann and G. Endrodi for their lattice data.
N. N. S. also would like to thank P. Allen for useful discussions.


\begin{thebibliography}{99}

\bibitem{duncan}
R. C. Duncan and C. Thompson,
Astrophys. J. {\bf 392}, L9 (1992);
C. Kouveliotou et al.,
Nature {\bf 393}, 235 (1998).

\bibitem{HIC}
V. Skokov, A. Y. Illarionov, and V. Toneev,
Int. J. Mod. Phys. A {\bf 24}, 5925 (2009);
V. Voronyuk, V. Toneev, W. Cassing, E. Bratkovskaya, V. Konchakovski, and S. Voloshin,
Phys. Rev. C {\bf 83}, 054911 (2011).

\bibitem{kharzeev} 
D.~E.~Kharzeev, L.~D.~McLerran and H.~J.~Warringa,
Nucl.\ Phys.\ A {\bf 803}, 227 (2008).

\bibitem{cosmo}
T. Vachaspati,
Phys. Lett. B {\bf265}, 258 (1991);
K. Enqvist and P. Olesen,
Phys. Lett. B {\bf319}, 178 (1993).

\bibitem{Klimenko:1991he} 
K.~G.~Klimenko,
Z.\ Phys.\ C {\bf 54}, 323 (1992).

\bibitem{Ebert:2003yk} 
D.~Ebert and K.~G.~Klimenko,
Nucl.\ Phys.\ A {\bf 728}, 203 (2003).

\bibitem{Ferrer:2005vd} 
E.~J.~Ferrer, V.~de la Incera and C.~Manuel,
Phys.\ Rev.\ Lett.\  {\bf 95}, 152002 (2005).

\bibitem{Mizher:2010zb}
A.~J.~Mizher, M.~N.~Chernodub and E.~S.~Fraga,
Phys.\ Rev.\ D {\bf 82}, 105016 (2010).

\bibitem{Chatterjee}
B. Chatterjee, H. Mishra and A. Mishra,
Phys. Rev. D {\bf 84}, 014016 (2011).

\bibitem{Chernodub:2011mc} 
M.~N.~Chernodub,
Phys.\ Rev.\ Lett.\  {\bf 106}, 142003 (2011).

\bibitem{baliJHEP2012}
G. S. Bali, F. Bruckmann, G. Endr\"odi, Z. Fodor, S. D. Katz, S. Krieg, 
A. Sch\"afer, and K. K. Szab\'o,
J. High Energy Phys. {\bf 1202}, 044 (2012).

\bibitem{DElia3}
M. D'Elia,
Lect.\ Notes Phys.\ {\bf 871}, 181 (2013).

\bibitem{ruggieri2}
R.~Gatto and M.~Ruggieri,
Lect.\ Notes Phys.\  {\bf 871}, 87 (2013).

\bibitem{Fraga:2012rr} 
E.~S.~Fraga,
Lect.\ Notes Phys.\  {\bf 871}, 121 (2013).

\bibitem{Klevansky}
S. P. Klevansky and R. H. Lemmer, 
Phys. Rev. D {\bf 39}, 3478 (1989).

\bibitem{avancini2012}
Sidney S. Avancini, D\'{e}bora P. Menezes, Marcus B. Pinto, and Constan\c{c}a Provid\^{e}ncia,
Phys. Rev. D (R) {\bf 85}, 091901 (2012).

\bibitem{DElia1}
M. D'Elia, S. Mukherjee, and F. Sanfilippo, 
Phys. Rev. D {\bf 82}, 051501 (2010).

\bibitem{DElia2}
M. D'Elia and F. Negro, 
Phys. Rev. D {\bf 83}, 114028 (2011).

\bibitem{Ilgenfritz1}
E.-M. Ilgenfritz, M. Kalinowski, M. M\"uller-Preussker, B. Petersson, and A. Schreiber,
Phys. Rev. D {\bf 85}, 114504 (2012).

\bibitem{Ilgenfritz2}
E.-M. Ilgenfritz, M. M\"uller-Preussker, B. Petersson, and A. Schreiber,
arXiv:1310.7876v1  [hep-lat].

\bibitem{bali2012}
G. S. Bali, F. Bruckmann, G. Endr\"odi, Z. Fodor, S. D. Katz, and A. Sch\"afer,
Phys. Rev. D {\bf 86}, 071502 (2012).

\bibitem{endrodi2013} 
F. Bruckmann, G. Endrodi and T. G. Kovacs,
J. High Energy Phys. {\bf 1304}, 112 (2013).

\bibitem{PNJL}
K. Fukushima, Phys. Lett. B {\bf591}, 277 (2004);
C. Ratti, M. A. Thaler, and W. Weise,
Phys. Rev. D {\bf 73}, 014019 (2006);
E.~Megias, E.~Ruiz Arriola and L.~L.~Salcedo,
 Phys.\ Rev.\  D {\bf 74}, 065005 (2006).

\bibitem{EPNJL}
Y. Sakai, T. Sasaki, H. Kouno, and M. Yahiro,
Phys. Rev. D {\bf 82}, 076003 (2010);
J. Phys. G {\bf 39}, 035004 (2012).
T. Sasaki, Y. Sakai, H. Kouno, and M. Yahiro,
Phys. Rev. D {\bf 84}, 091901(R) (2011).

\bibitem{ruggieri}
K. Fukushima, M. Ruggieri and R. Gatto,
Phys. Rev. D {\bf 81}, 114031 (2010);

\bibitem{njlsu3}
T. Hatsuda and T. Kunihiro,
Phys. Rep. {\bf 247}, 221 (1994);
S.P. Klevansky,
Rev.\ Mod.\ Phys.\  {\bf 64}, 649 (1992).

\bibitem{Buballa:2003qv}
M. Buballa, Phys. Rep. {\bf 407}, 205 (2005).

\bibitem{Ratti:2006}
S.~Roessner, C.~Ratti and W.~Weise,
Phys.\ Rev.\ D {\bf 75}, 034007 (2007).

\bibitem{Klev_param}
P. Rehberg, S.P. Klevansky, and J. H\"ufner,
Phys. Rev. C {\bf 53}, 410 (1996).

\bibitem{prc}
D.P. Menezes, M.B. Pinto, S.S. Avancini, A. P\'{e}rez Mart\'{\i}nez and C. Provid\^{e}ncia,
Phys. Rev. C {\bf 79}, 035807 (2009);
D.P. Menezes, M.B. Pinto, S.S. Avancini and C. Provid\^{e}ncia,
Phys. Rev. C {\bf 80}, 065805 (2009).

\bibitem{hotnjl}
S.S. Avancini, D.P. Menezes and C. Provid\^encia,
Phys. Rev. C {\bf 83}, 065805 (2011).

\bibitem{costa}
P. Costa, M.C. Ruivo, C.A. de Sousa and H. Hansen,
Symmetry 2(3), 1338 (2010);
P.~Costa, M.~C.~Ruivo, C.~A.~de Sousa, H.~Hansen and W.~M.~Alberico,
Phys.\ Rev.\ D {\bf 79}, 116003 (2009).

\bibitem{ruggieri3}
R. Gatto and M. Ruggieri,
Phys. Rev. D {\bf 82}, 054027 (2010);
Phys. Rev. D {\bf 83}, 034016.

\bibitem{Skokov}
V. Skokov,
Phys. Rev. D {\bf 85}, 034026 (2012).

\bibitem{Nam}
Seung-il Nam and Chung-Wen Kao,
Phys. Rev. D {\bf 83}, 096009 (2011).

\bibitem{Miransky}
V. A. Miransky and I. A. Shovkovy, 
Phys. Rev. D {\bf 66}, 045006 (2002).

\bibitem{Kojo:2012js} 
T.~Kojo and N.~Su,
Phys.\ Lett.\ B {\bf 720}, 192 (2013).

\bibitem{wambach}  
B.~-J.~Schaefer, J.~M.~Pawlowski and J.~Wambach,
Phys.\ Rev.\ D {\bf 76 }, 074023  (2007);
B.~-J.~Schaefer, M.~Wagner and J.~Wambach,
Phys.\ Rev.\  D {\bf 81 }, 074013 (2010);
T.~K.~Herbst, J.~M.~Pawlowski and B.~J.~Schaefer,
Phys.\ Lett.\  B {\bf 696}, 58 (2011).

\bibitem{Pawlowski2011}
T. K. Herbst, J. M. Pawlowski and B.-J. Schaefer,
Phys. Lett. B {\bf 696}, 58 (2011).

\bibitem{Fukushima:2012kc} 
K.~Fukushima and Y.~Hidaka,
Phys.\ Rev.\ Lett.\  {\bf 110}, 031601 (2013).

\bibitem{Chao:2013qpa} 
J.~Chao, P.~Chu and M.~Huang,
Phys.\ Rev.\ D {\bf 88}, 054009 (2013).


\end{thebibliography}
\end{document}